\documentclass[conference]{IEEEtran}
\IEEEoverridecommandlockouts
\usepackage{cite}
\usepackage{amsmath,amssymb,amsfonts}
\usepackage{algorithmic}
\usepackage{algorithm}
\usepackage{graphicx}
\usepackage{subcaption}
\usepackage{textcomp}
\usepackage{xcolor}
\usepackage{dsfont}
\usepackage{bm}
\usepackage{siunitx}
\usepackage{url}
\usepackage{hyperref}

\newcommand{\figref}[1]{Fig.\,\ref{#1}}
\newcommand{\Figref}[1]{Figure\,\ref{#1}}

\newcommand{\secref}[1]{Sec.\,\ref{#1}}

\newcommand{\mE}{{\mathbb{E}}}
\newcommand{\mP}{{\mathbb{P}}}

\newcommand{\R}{\mathbb{R}}

\usepackage{amsthm}

\newtheorem{assumption}{Assumption}
\newtheorem{remark}{Remark}

\begin{document}

\title{Combined Plant and Control Co-design via Solutions of Hamilton-Jacobi-Bellman Equation Based on Physics-informed Learning\\
}

\author{
\IEEEauthorblockN{Kenjiro Nishimura,~~ Hikaru Hoshino,~~ Eiko Furutani}
\IEEEauthorblockA{\textit{Department of Electrical Materials and Engineering} \\
\textit{University of Hyogo}\\
2167 Shosya, Himeji, Hyogo 671-2280, Japan \\
er24h020@guh.u-hyogo.ac.jp, \{hoshino, furutani\}@eng.u-hyogo.ac.jp }
% \and
% \IEEEauthorblockN{Hikaru Hoshino}
% \IEEEauthorblockA{\textit{Department of Electrical Materials and Engineering} \\
% \textit{University of Hyogo}\\
% 2167 Shosya,Himeji,Hyogo 671-2280, Japan \\
% hoshino@eng.u-hyogo.ac.jp}
% \and
% \IEaEauthorblockN{Eiko Furutani}
% \IEEEauthorblockA{\textit{Depertment of Electrical Materials and Engineering} \\
% \textit{University of Hyogo}\\
% 2167 Shosya,Himeji,Hyogo 671-2280, Japan \\
% furutani@eng.u-hyogo.ac.jp}
}

\maketitle

\begin{abstract}
This paper addresses integrated design of engineering systems, where physical structure of the plant and controller design are optimized simultaneously. To cope with uncertainties due to noises acting on the dynamics and modeling errors, an Uncertain Control Co-design (UCCD) problem formulation is proposed. Existing UCCD methods usually rely on uncertainty propagation analyses using Monte Calro methods for open-loop solutions of optimal control, which suffer from stringent trade-offs among accuracy, time horizon, and computational time. The proposed method utilizes closed-loop solutions characterized by the Hamilton-Jacobi-Bellman equation, a Partial Differential Equation (PDE) defined on the state space. A solution algorithm for the proposed UCCD formulation is developed based on PDE solutions of Physics-informed Neural Networks (PINNs). Numerical examples of regulator design problems are provided, and it is shown that simultaneous update of PINN weights and the design parameters effectively works for solving UCCD problems. 
\end{abstract}


%%%%%%%%%%%%%%%%%%%%%%%%%%%%%%%%%%%%%%%%%%%%%%%%%%%%%%%%%%%%%
\section{Introduction} \label{sec:introduction}

When designing an engineering system, one can adopt a sequential strategy where the physical system design is optimized first, followed by the controller design. 
For instance, consider the system-level design of a robotic manipulator.
The physical system design may optimize the geometric parameters of the links, and the control design may determine the joint torque time trajectories for specific tasks~\cite{Allison2013}.
Although the physical system design is often performed based only on static characteristics of the system, the resulting systems are suboptimal in most cases, and it is desirable to consider dynamic characteristics in the physical system design in cooperation with the controller design.  
This view is called as Control Co-design (CCD) approach \cite{Garcia-Sanz2019}, and many authors have shown its benefit in various applications including robotic manipulators~\cite{Allison2013, Fadini2022}, quadruped robots~\cite{Fadini2024}, flexible space structures~\cite{Onoda1987,Chilan2017}, electric motors~\cite{Reyer2002}, offshore wind farms~\cite{Sharma2024}, Field Programmable Gate Array (FPGA) circuits~\cite{Dang2019}, and civil structures~\cite{Alavi2021}. 

One of the main challenges in CCD that remains to be addressed is to consider the impact of uncertainties coming from the noise acting on the control channels, unmodeled or neglected dynamics of the system, estimation errors in model parameters, and so on. 
All of these uncertainties may propagate through the dynamical system and transform the states into uncertain trajectories \cite{Azad2023}.
These problems are termed as Uncertain CCD (UCCD)~\cite{Azad2023} or Robust CCD~\cite{Bravo-Palacios2021,Fadini2022}, and several formulations and solution methods have been proposed. 
%In \cite{Azad2023}, three general formulations for UCCD problems are explored motivated by concepts from stochastic programming, robust optimization, and fuzzy programming. 
In \cite{Giordano2018,Brault2021}, it is proposed to optimize a metric that represents the sensitivity of the trajectory to perturbation.
This can reduce the sensitivity of the trajectory for a specific uncertainty relying on custom-made cost formulations, but it increases the complexity of the problem and the cost function has to be carefully designed.  
Another approach is stochastic programming, in which the optimal trajectory is found for a set of perturbed scenarios \cite{Bravo-Palacios2021}.
In \cite{Fadini2022,Fadini2024}, a bi-level optimization scheme, where trajectories are optimized by the Differential Dynamic Programming (DDP) algorithm in an inner loop, and hardware is optimized in an outer loop with a genetic algorithm based on Covariance Matrix Adaptation Evolution Strategy (CMA-ES)~\cite{Hansen2016}.
However, the above methods rely on Monte Calro simulation for analyzing uncertainty propagation through dynamical systems, and are subject to stringent trade-offs among accuracy, time horizon, and computational time~\cite{Zhang2021}.

In this paper, we propose a novel approach to UCCD problems, where optimality of the trajectories are characterized by the Hamilton-Jacobi-Bellman (HJB) equation in optimal control. 
In contrast to DDP based approach in \cite{Fadini2022,Fadini2024}, where open-loop solutions of an optimal control problem are computed, the HJB equation characterizes closed-loop solutions in the form of a deterministic Partial Differential Equation (PDE), and uncertainty propagation can be naturally incorporated within the framework of stochastic control theory~\cite{Yong1999,Fleming06}. 
In our previous work~\cite{Hoshino2023}, we proposed a CCD solution method for deterministic dynamics with uncertainties in initial conditions based on Galerkin-approximations of the HJB equation~\cite{Beard1997}. 
This paper presents an extension of the CCD framework in \cite{Hoshino2023} to deal with stochastic dynamics and uncertainties in model parameters. 
To the best of our knowledge, this is the first work to use the HJB equation for the purposes of solving UCCD problems. 
Furthermore, we present a UCCD solution method based on Physics-informed Neural Networks (PINNs)~\cite{Raissi2019}. 
PINNs are widely used for solving various PDEs~\cite{Cai2021,Cuomo2022} and have a potential in dealing with a high-dimensional system~\cite{Miao2022}, whereas the HJB equation is notoriously difficult to solve by using standard numerical methods when the dimension becomes 5 or more~\cite{Mitchell2008}. 
This paper presents numerical examples of regulator design for a 2-dimensional nonlinear system and Linear Quadratic Regulator (LQR) problems of up to 10 dimensional systems to examine the effectiveness of the proposed method. 

The rest of this paper is organized as follows. 
In \secref{sec:formulation}, the proposed UCCD problem formulation is presented. 
A solution method for the UCCD problem based PINNs is introduced in \secref{sec:method}. 
Numerical results are provided in \secref{sec:simulation}. 
Conclusions and future works are summarized in \secref{sec:conclusion}. 

%%%%%%%%%%%%%%%%%%%%%%%%%%%%%%
\subsection{Notation}

Let $\mathbb{R}$ be the set of real numbers, and $\R^n$ be the $n$-dimensional Euclidean space. %and the set of nonnegative real numbers, respectively. Let $\mZ$ and $\mZ_+$ be the set of integers and the set of non-negative integers. 
For an open set $A$, $\overline{A}$ stands for the closure of $A$, and $\partial A$ for the boundary of $A$. 
For a scalar function $\phi$, % with an argument $x \in\R^n$, 
$\partial_x \phi$ stands for the gradient of $\phi$ with respect to $x$, and $\partial_{x}^2 \phi$ for the Hessian matrix of $\phi$. 
Let $\mathrm{tr}(M)$ be the trace of the matrix $M$. 
%Let $\lfloor x \rfloor \in \mZ$ be the greatest integer less than or equal to $x\in\R$.  
%Let $\mathds{1}[\mathcal{E}]$ be an indicator function, which takes 1 when the condition $\mathcal{E}$ holds and otherwise 0.
%Let $\mP[ \mathcal{E} | X_0 = x ]$ represents the probability that the condition $\mathcal{E}$ holds involving a stochastic process $X = \{ X_t \}_{t\in\R_+}$ conditioned on $X_0 = x$.  
For random variables $X$ and $Y$, let $\mathbb{E}[X]$ be the expectation of $X$, and $\mathbb{E}[X|Y=y]$ be the conditional expectation of $X$ given $Y=y$. 
We use upper-case letters (\emph{e.g.}, $Y$) to denote random variables and lower-case letters (\emph{e.g.}, $y$) to denote their specific realizations. 

%%%%%%%%%%%%%%%%%%%%%%%%%%%%%%%%%%%%%%%%%%%%%%%%%%%%%%%%%%%
\section{Co-design Problem Formulation} \label{sec:formulation}

In this section, we firstly introduce a basic co-design problem commonly studied in literature in \secref{sec:basic_formulation}, and then present the proposed formulation in \secref{sec:proposed_formulation}. 

%%%%%%%%%%%%%%%%%%%%%%%%%%%%%%%%
\subsection{Deterministic Control Co-design Problem} \label{sec:basic_formulation}

There are two commonly studied strategies for CCD problems: simultaneous and nested \cite{Herber2019}.
The simultaneous strategy optimizes both the plant and control variables in a same optimization formulation and is the most fundamental representation \cite{Herber2019}. 
The nested strategy can be seen as a specific reorganization of the simultaneous formulation. An outer loop optimizes the design of the controlled plant, and an inner loop identifies the optimal control for each plant design tested by the outer loop. 
Here we briefly introduce a basic formulation in the simultaneous strategy. 
Consider the following dynamical system with an $l$-dimensional design parameter $\rho=[\rho_1, \, \dots, \, \rho_l ]^\top$: 
\begin{align} 
 \dot{x} = f( x, u; \rho )  \label{eq:system}
\end{align}
where $x \in \mathbb{R}^n$ stands for the $n$-dimensional state, $\dot{x}$ for the time derivative of $x$, and $u \in \mathbb{R}^m $ for the $m$-dimensional control action.  
For this system, a standard optimal control problem can be considered with the following cost integral $J_\mathrm{c}$: 
\begin{align}
 J_\mathrm{c}(\rho, u(\cdot)) =  
 \int_0^T  L ( x(t), u(t), \rho) \mathrm{d}t 
 + M(x(T), \rho)
 \label{eq:cost_integral}
\end{align}
where $T$ stands for the horizon length, and $L$ and $M$ for the Lagrange cost (running cost) and Mayer cost (terminal cost), respectively.  
The above notation shows that the cost integral $J_\mathrm{c}$ can be affected by the design parameter $\rho$ through the dependence of $L$ and $M$ on $\rho$. 
Besides, $J_\mathrm{c}$ depends on $\rho$ through the change in the dynamics \eqref{eq:system}. 
By combining the cost $J_\mathrm{c}$ representing the control performance and an additional cost $J_\mathrm{p}$ about the choice of the parameter $\rho$ of the plant, which represents, \emph{e.g.}, hardware materials costs or assembling costs, the co-design problem can be formulated as 
\begin{align}
 &\min_{ \rho,\, u(\cdot) } \quad w_\mathrm{p} J_\mathrm{p}(\rho) + w_\mathrm{c} J_\mathrm{c}(\rho,\,u(\cdot))  \notag \\
 & \mathrm{s.t.} \quad  \dot{x}(t) = f(x(t), u(t); \rho), \quad \forall t\in [0, T],  \notag \\
 &\hspace{8mm}  x(0) = x_\mathrm{0}  
\end{align}
where $w_\mathrm{p}$ and $w_\mathrm{c}$ are weighting coefficients, and $x_0$ stands for the initial state of the optimal control problem. 
Although a more concise formulation can be obtained by including the term $J_\mathrm{p}$ in the Mayer term of the optimal control problem, the above formulation is commonly used to allow for more natural representations of CCD problems.

%%%%%%%%%%%%%%%%%%%%%%%%%%%%%%%%%
\subsection{Proposed Formulation} \label{sec:proposed_formulation}

Here we present the proposed formulation for UCCD problems. 
Let $(\Omega, \mathcal{F}, \{\mathcal{F}_t\}_{t\ge0}, \mP)$ be a filtered probability space, and consider a control system with stochastic noise represented by $\mathcal{F}_t$-standard $w$-dimensional Brownian motion $\{ W_t \}_{t \ge 0}$ starting from $W_0 = 0$.
For an open set $\mathbb{X} \subset \R^n$ in the $n$-dimensional state space, the state $X_t \in \overline{\mathbb{X}}$ evolves according to the following Stochastic Differential Equation (SDE): 
\begin{align}
 \mathrm{d}X_t = \{ f(X_t;\rho)+g(X_t;\rho)U_t \}\mathrm{d}t + \sigma(X_t; \rho) \mathrm{d}W_t,  \label{eq:sde}   
\end{align}
where $\rho$ stands for the design parameter of the hardware, and $\{U_t\}_{t\ge0}$ with $U_t \in \mathbb{U} \subset \mathbb{R}^m$ is 
an $\{\mathcal{F}_t\}_{t\ge0}$-adapted control process, which takes values in $\mathbb{U}$. 
Throughout this paper, we assume sufficient regularity in the coefficients of the system \eqref{eq:sde}.
That is, the functions $f$, $g$ and $\sigma$ are chosen in a way such that the SDE \eqref{eq:sde} admits a unique strong solution (see, e.g., Section IV.2 of \cite{Fleming06}). 
The size of $\sigma(X_t; \rho)$ is determined from the uncertainties in the disturbance, unmodeled dynamics, and prediction errors of the environmental variables. 
As we solve the equation in the domain $\mathbb{X}$, define a stopping time $\tau$ as 
\begin{align}
    \tau = \inf\{ t ~| ~ X_t \notin \mathbb{X} \}. 
\end{align}
Then, for each $\rho$, consider an optimal control problem to minimize the following cost functional: 
\begin{align} \label{eq:cost_term_Jc}
    J_\mathrm{c}(\rho, x, U) =  & \mE \biggl[ \int_0^\tau L(X_s, U_s, \rho) \mathrm{e}^{-\gamma s} \mathrm{d}s \notag \\ 
    & \hspace{8mm} + \mathrm{e}^{-\gamma \tau} M(X_\tau, \rho) ~|~ X_0 = x  \biggr],
\end{align}
where $L$ and $M$ represent Lagrange and Meyer costs as mentioned in \secref{sec:basic_formulation}, and $\gamma$ stands for the discount factor. 
The control process $U:=\{U_t\}_{t\ge0}$ is chosen over a set $\mathcal{U}$ of admissible control processes that have values in $\mathbb{U}$ and are adapted to the filtration $\{\mathcal{F}_t\}_{t\ge0}$. 
Then, by defining the value function $V$ as 
\begin{align} \label{eq:value_function}
    V(\rho, x) = \inf_{U\in\mathcal{U}} J_\mathrm{c} (\rho, x, U), 
\end{align}
the control performance is characterized as a function of the parameter $\rho$ and the initial state $x$. 
From stochastic control theory~\cite{Yong1999,Fleming06}, it can be shown that the value function \eqref{eq:value_function} satisfies the following HJB equation:
% HJB
\begin{align}
 \inf_{u \in \mathbb{U}} \left\{ \mathcal{L}^u V(\rho, x) + L(x,u,\rho) -\gamma V(\rho, x) \right\}=0,   \label{eq:hjb_equation}
\end{align}
in $\mathbb{X}$, where $\mathcal{L}^u$ is defined by 
\begin{align}
    \mathcal{L}^u V(\rho,x) := 
    & \dfrac{1}{2} \mathrm{Tr} \left[ \sigma^\top \sigma (\partial_x^2 V) \right] (\rho, x) \notag \\
    & + \{ f(x;\rho)+g(x;\rho)u\}^\top \partial_x V(\rho,x)
\end{align}
with the boundary condition 
\begin{align} \label{eq:boundary_condition}
 V(\rho, x) = M(\rho, x), 
\end{align}
on $\partial \mathbb{X}$. 
By using the above, the UCCD problem proposed in this paper is formulated as
\begin{align} \label{eq:UCCD}
 &\min_{ \rho }~ J = w_\mathrm{p}  J_\mathrm{p}(\rho)  + w_\mathrm{c} \int_\mathbb{X} \omega(x) V(\rho, \, x) \mathrm{d}x  \notag \\
  & \hspace{1mm} \mathrm{s.t.}  \notag \\
  & \hspace{3mm} \inf_{u \in \mathbb{U}} \left\{ \mathcal{L}^u V(\rho, x) + L(x,u,\rho) -\gamma V(\rho, x) \right\}=0, \quad x \in \mathbb{X},   \notag \\
  & \hspace{4mm} V(\rho, x) = M(\rho, x), \quad x \in \partial \mathbb{X}
\end{align}
where $\omega: \mathbb{R}^n \to \mathbb{R}$ is a weighting function to take an expectation of the value $V(\rho, x)$ satisfying
\begin{align}
 \int_\mathbb{X} \omega(x) \mathrm{d}x = 1. 
 \end{align}
The problem \eqref{eq:UCCD} is a partial differential equation constrained optimization problem, and a proper computational technique is needed to be solved.

\begin{remark}
 In the above formulation, both the effects of the noise acting on the control channels and unmodeled/neglected dynamics are represented by the nonlinear function $\sigma$. 
 Also, uncertainties in initial conditions are represented by the distribution $\omega$. Furthermore, if there are uncertain model parameters, denoted by $\phi$, the system dynamics can be augmented as 
 \begin{align} \label{eq:augmented_system}
   \begin{bmatrix} \mathrm{d}X_t \\ \mathrm{d}\phi \end{bmatrix}= 
   & \begin{bmatrix} f(X_t;\rho, \phi)+g(X_t; \rho, \phi) U_t \\ 0 \end{bmatrix} \mathrm{d}t
   \notag \\ 
   & \hspace{5mm} +\begin{bmatrix} \sigma(X_t; \rho) \\ 0 \end{bmatrix} \mathrm{d}W_t.        
 \end{align}
 Thus, by applying the formulation \eqref{eq:UCCD} adopted for the augmented system \eqref{eq:augmented_system}, uncertainties in model parameters can also be captured by the distribution $\omega$. 
 In conclusion, the proposed formulation can capture all the uncertainties due to the noise acting on the control channels, unmodeled/neglected dynamics, initial conditions, and model parameters. 
\end{remark}

%%%%%%%%%%%%%%%%%%%%%%%%%%%%%%%%%%%%%%%%%%%%%%%%%%%%%%%%%%%
\section{Solution algorithm} \label{sec:method}

%%%%%%%%%%%%%%%%%%%%%%%%%%%%%%%%%%
% Fig: Solusion overview 
%%%%%%%%%%%%%%%%%%%%%%%%%%%%%%%%%%
\begin{figure}[!t]
    \centering
    \includegraphics[width=0.95\linewidth]{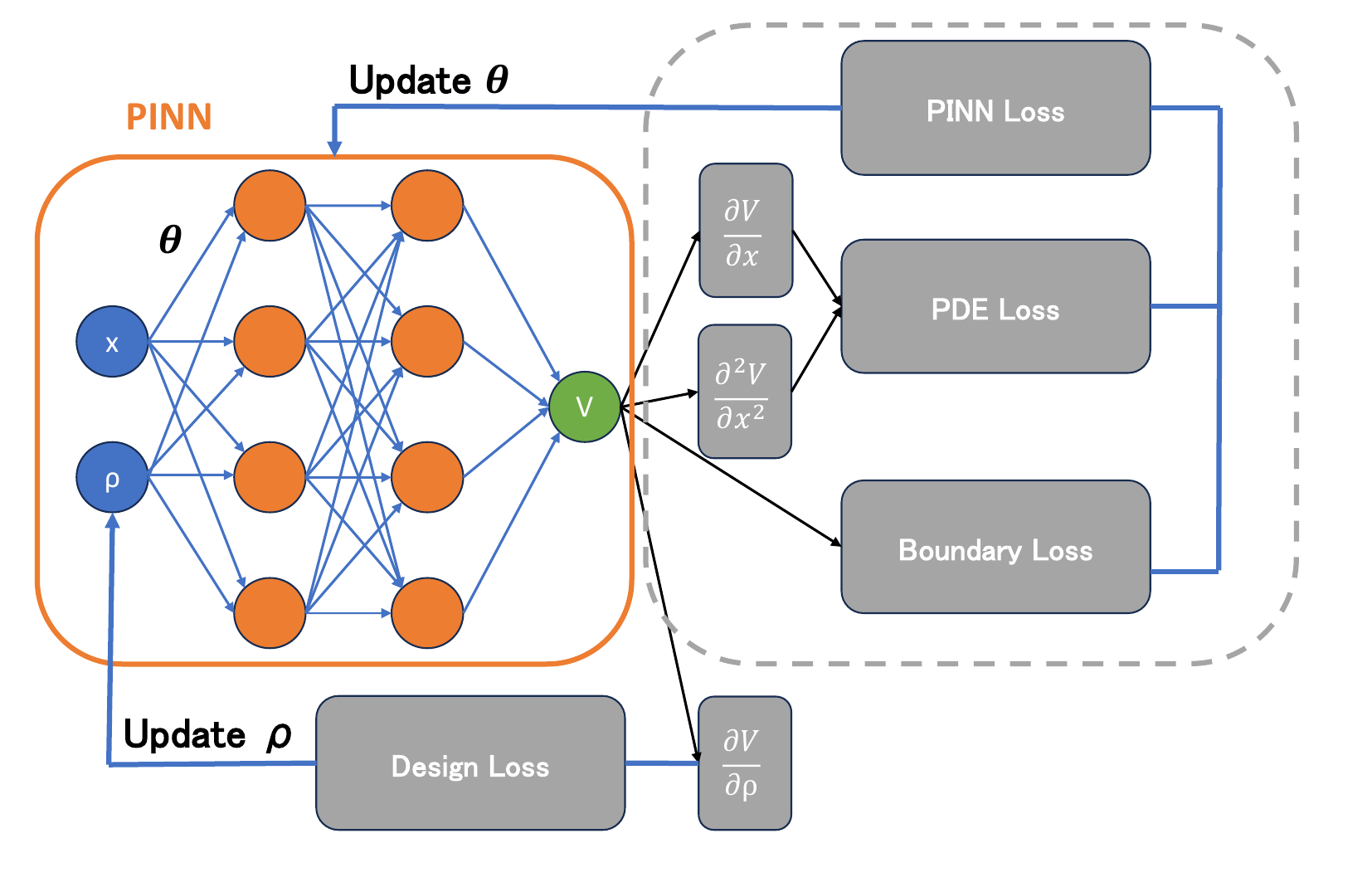}
    \caption{Schematic diagram of the proposed UCCD method}
    \label{fig:method}
\end{figure}

This section presents a solution method for the proposed formulation for UCCD problems. 
To deal with the HJB equation as a PDE constraint in the optimization problem \eqref{eq:UCCD}, we use the PINN framework, which is able to solve PDEs by exploiting machine learning techniques~\cite{Raissi2019}.  
\Figref{fig:method} shows a schematic overview of the proposed UCCD method based on PINN. 
The PINN takes the pair $(\rho, x)$ of the design parameter $\rho$ and the state $x$ (as well as the uncertain model parameter $\phi$ stated in Remark\,1 if the augmented representation is used), and outputs the prediction $\widehat{V}$ of the value $V(\rho, x)$, as well as its derivatives $\partial \widehat{V}/\partial \rho$ and $\partial \widehat{V}/\partial x$ computed by automatic differentiation~\cite{Baydin2018}. 
By assuming that the PINN is parameterized by $\theta$, the loss function $L_\mathrm{PINN}$ for the learning is defined as \begin{align} \label{eq:loss_pinn}
    L_\mathrm{PINN}(\theta, \mathcal{S}_\mathrm{h}, \mathcal{S}_\mathrm{b}) = \mu_\mathrm{h} L_\mathrm{HJB}(\theta, \mathcal{S}_\mathrm{h}) + \mu_\mathrm{b} L_\mathrm{bdry}(\theta,\mathcal{S}_\mathrm{b}) 
\end{align}
where $L_\mathrm{HJB}$ and $L_\mathrm{bdry}$ are loss terms for the HJB equation \eqref{eq:hjb_equation} and the boundary condition \eqref{eq:boundary_condition}, respectively, and $\mu_\mathrm{h}$ and $\mu_\mathrm{b}$ are the weighting coefficients.
The HJB loss term $L_\mathrm{HJB}$ is a function of $\theta$ and a set of $N_\mathrm{h}$ random samples with $\mathcal{S}_\mathrm{h} = \{ (\rho_i, x_i) \,|\, i\in \{1,\dots N_\mathrm{h} \}, x_i \in \mathbb{X} \}$, and  given by 
\begin{align} \label{eq:loss_hjb}
    L_\mathrm{HJB}(\theta, \mathcal{S}_\mathrm{h}) = \dfrac{1}{N_\mathrm{p}} \sum_{i=1}^{N_\mathrm{p}} \|F(\rho_i, x_i, \theta) \|  ^2 
\end{align}
with 
 \begin{align}
  F(\rho, x, \theta) &= \inf_{u \in \mathbb{U}} \left\{ \mathcal{L}^u \widehat{V}(\rho, x, \theta) + L(x,u,\rho) -\gamma \widehat{V}(\rho, x) \right\}. 
\end{align}
Here, the optimal control $u$ needs to be addressed, and in this paper, we assume a specific form of the cost function. 
\begin{assumption}
    The Lagrange cost term $L$ in the cost functional $J_\mathrm{c}$ in \eqref{eq:cost_term_Jc} takes the following form  
    \begin{align}
        L(x, u) = \hat{L}(x) + u^\top \mathsf{R} u, 
    \end{align}
    where $\mathsf{R}$ is a positive-definite matrix. 
\end{assumption}
\noindent 
With this assumption, we have an explicit optimal control as 
\begin{align}
  u^\ast(x) = -\dfrac{1}{2} \mathsf{R}^{-1} g(x)^\top  \partial_x V. 
\end{align}
%Extension of our solution algorithm to problems with implicit optimal control is in our future work. 
The boundary loss term $L_\mathrm{bdry}$ in \eqref{eq:loss_pinn} is computed from a set of $N_\mathrm{b}$ random samples with $\mathcal{S}_\mathrm{b} = \{ (\rho_j, x_j) \,|\, j\in \{1,\dots N_\mathrm{b} \}, x_j \in \partial\mathbb{X} \}$, and given by 
\begin{align} \label{eq:loss_boudary}
 L_\mathrm{bdry}(\theta,\mathcal{S}_\mathrm{b}) = \dfrac{1}{N_\mathrm{b}} \sum_{j=1}^{N_\mathrm{b}} \| \widehat{V}(\rho_j, x_j, \theta) - M(\rho_j, x_j) \|^2. 
\end{align}

To solve the UCCD problem \eqref{eq:UCCD}, we propose to simultaneously update the weights $\theta$ of the PINN and the design parameter $\rho$. 
The proposed solution method is presented in Algorithm\,\ref{alg:UCCD}. 
At each epoch, the sets $\mathcal{S}_\mathrm{h} = \{ (\rho_i, x_i )\}_{i=1}^{N_\mathrm{h}}$ and $\mathcal{S}_\mathrm{b} = \{ (\rho_j, x_j )\}_{j=1}^{N_\mathrm{b}}$ are randomly sampled with $x_i \in \mathbb{X}$ and $x_j \in \partial\mathbb{X}$. 
The samples $\rho_i$ and $\rho_j$ are drawn by 
\begin{align}
    \rho_i=\rho+\epsilon_i, \quad \rho_j = \rho + \epsilon_j, \quad \epsilon_i,\epsilon_j \sim \mathcal{N}(0, \sigma_\mathrm{n}^2)
\end{align}
where $\mathcal{N}(0, \sigma_\mathrm{n}^2)$ stands for the Gaussian distribution with the zero mean and the standard deviation of $\sigma_\mathrm{n}$. 
The noise term $\epsilon$ is added for exploration of $\rho$, and its effect is discussed in \secref{sec:simulation}. 
The minimization of the loss $L_\mathrm{PINN}$ imposes the constraints in the UCCD problem \eqref{eq:UCCD}. 
The design parameter $\rho$ is then updated to minimize the objective function $J$.  
To this end, a set $\mathcal{S}_\mathrm{r}$ of $N_\mathrm{r}$ random samples are drawn as $\mathcal{S}_\mathrm{r} = \{ x_k \,|,\, k\in \{ 1, \dots, N_\mathrm{r} \}, x_k \in \mathbb{X} \}$, and the following loss $L_\mathrm{r}$ is considered:
\begin{align} \label{eq:loss_rho}
    L_\mathrm{r}(\rho, \mathcal{S}_\mathrm{r}) =  w_\mathrm{p}  J_\mathrm{p}(\rho)  + w_\mathrm{c} \sum_{k=1}^{N_\mathrm{r}}  \widehat{V}(\rho, \, x_k)
\end{align}
Note that the samples $x_k$ needs to be drawn from the distribution $\omega$ in \eqref{eq:UCCD} to be unbiased. 
The parameter $\rho$ is updated by performing a gradient step on the loss $ L_\mathrm{r} $ once in $N_\mathrm{up}$ epochs. 

%%%%%%%%%%%%%%%%%%%%%%%%%%%%%%%%%%%
% Algorithm 
%%%%%%%%%%%%%%%%%%%%%%%%%%%%%%%%%%%
\begin{figure}[!t]
  \begin{algorithm}[H]
      \caption{UCCD by solutions of HJB using PINN}
      \label{alg:UCCD}
      \begin{algorithmic}[1]
      \STATE Initialize the PINN $\widehat{V}$ with weights $\theta$
      \STATE Initialize the design parameter $\rho$ 
      \FOR{epoch = 1:M}
      \STATE Sample $\mathcal{S}_\mathrm{h} = \{(\rho_i,x_i) \}$ with $\rho_i=\rho+\epsilon_i$, $x_i \in \mathbb{X}$     
      \STATE Calculate the loss term $L_\mathrm{HJB}$ in \eqref{eq:loss_hjb}
      \STATE Sample $\mathcal{S}_\mathrm{b} = \{(\rho_j,x_j) \}$ with $\rho_j=\rho+\epsilon_j$, $x_j \in \partial\mathbb{X}$       
      \STATE Calculate the loss term $L_\mathrm{bdry}$ in \eqref{eq:loss_boudary}
      \STATE Perform a gradient descent step on the loss $L_\mathrm{PINN}$ in \eqref{eq:loss_pinn}
      with respect to $\theta$
      \IF {epoch\,\%\,$N_\mathrm{up}$ == 0 }
      \STATE Sample $\mathcal{S}_\mathrm{r} = \{ x_k \}$ with $x_k \sim \omega$
      \STATE Perform a gradient descent step on the loss $L_\mathrm{r}$ in \eqref{eq:loss_rho} with respect to $\rho$
      \ENDIF
      \ENDFOR
      \end{algorithmic}
  \end{algorithm}
\end{figure}

%%%%%%%%%%%%%%%%%%%%%%%%%%%%%%%%%%%%%%%%%%%%%%%%%%%%%%%
\section{Numerical Examples} \label{sec:simulation}

This section provides numerical examples of CCD problems for regulator designs.
A 2-dimensional nonlinear deterministic dynamical system is treated in \secref{sec:planer}, and LQR problems for up to 10-dimensional systems are treated in \secref{sec:stochasticLQR}. 
The algorithm was implemented by a deep learning framework PyTorch \cite{Paszke2019}, and our implementation is available at \url{https://github.com/er24h020/SCIS_ISIS_2024}. 

%%%%%%%%%%%%%%%%%
\subsection{Deterministic Nonlinear Planar System} \label{sec:planer}

We consider a CCD problem for a planer nonlinear dynamical system with uncertainty in initial states, for which a numerical analysis has been performed in our previous work~\cite{Hoshino2023} based on a Galerkin approximation-based CCD method.  
Consider the following deterministic dynamics:
\begin{align}
 \underbrace{\begin{bmatrix} \dot{x}_1 \\ \dot{x}_2 \end{bmatrix}}_{\dot{x} } 
 = 
 \underbrace{
 \begin{bmatrix} -x_1^3 -x_2  \\ x_1 +x_2 \end{bmatrix}
 }_{f(x; \rho)}
 + \underbrace{
 \begin{bmatrix} 0 \\ \rho \end{bmatrix} 
 }_{g(x; \rho)}
 u  \label{eq:target_system}
\end{align}
where $x = [x_1, x_2]^\top \in \R^2 $ stands for the state, $u \in  \R$ for the input, and $\rho\in \mathbb{R}$ for the design parameter. 
The objective function for the optimal control is given by 
\begin{align}
 J_\mathrm{c}(\rho, x) = \int_0^\infty p x(t)^\top x(t) + q u^2(t) \mathrm{d}t,
\end{align}
with $p=q=1$.
With this cost function, we have the following HJB equation:
\begin{align}
 \partial_x V^\top (f + g u)  +  p x^\top x+ q u^2 =0 
\end{align}
with $u= -\rho \partial_{x_2}V/(2q)$. 
The boundary condition is given as $V(\rho, 0) = 0$. 
Then, consider the CCD problem with the following objective function:
\begin{align}
 J = w_\mathrm{p} |\rho| + w_\mathrm{c} \int_\mathbb{X} \omega(\bm{x}) V(\bm{x}) \mathrm{d}x 
\end{align}
where $\mathbb{X} =  \left\{ (x_1, x_2) \in \mathbb{R}^2 ~\middle|~ |x_1|\le 1, |x_2|\le 1  \right\}$, $w_\mathrm{p}=1$,  $w_\mathrm{c}=4$, and $\omega(x)\equiv 1/4$. 
With this problem setting, the control performance is improved as $\rho$ increases, but a penalty is imposed on the increase in $\rho$ by the first term of the objective function. 

The objective function $J$ was minimized by Algorithm\,\ref{alg:UCCD}. 
For the function approximator $\widehat{V}$ of the value function, we used a neural network with 3 hidden layers with 64 units per layer, and the hyperbolic tangent (\texttt{tanh}) as the activation function. 
The soft plus function was used at the output. 
For updating the neural network's weights, \texttt{Adam} optimizer was used with the learning rate of $\SI{3E-3}{}$. 
The weights in the loss \eqref{eq:loss_pinn} was set as $\mu_\mathrm{h}=\mu_\mathrm{b}=1$, and the number of samples was $N_\mathrm{p}=1000$ for the inside of $\mathbb{X}$ and $N_\mathrm{b}=100$ for the boundary. 
The design parameter $\rho$ was initially set to $\rho = 1$, and updated by using \texttt{Adam} optimizer with the learning rate of $\SI{2e-2}{}$ at every 1 or 500 epochs ($N_\mathrm{up}=1$ or $500$) with $N_\mathrm{r}=1000$. 

%%%%%%%%%%%%%%%%%%%%%%%%%%%%%%%%%%%%%%
% Fig: NOLTA example
%%%%%%%%%%%%%%%%%%%%%%%%%%%%%%%%%%%%%%
\begin{figure}[!t]
    \centering
     \subcaptionbox{Change in design parameter $\rho$\label{fig:rho_nolta}}{\includegraphics[width=0.98\linewidth]{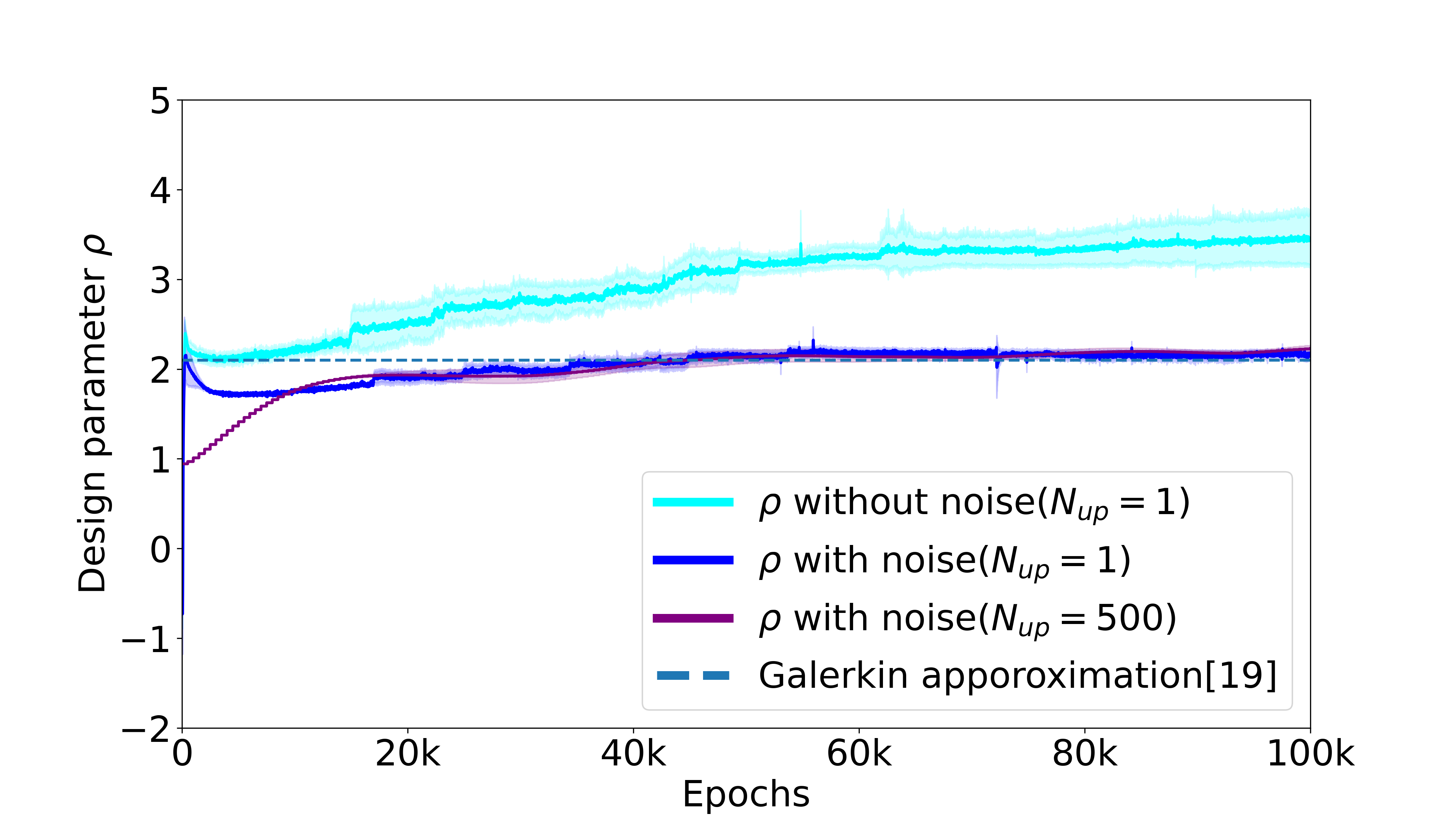}}\\[3mm]
     \subcaptionbox{Change in loss term $L_\mathrm{PINN}$\label{fig:loss_nolta}}
     {\includegraphics[width=0.98\linewidth]{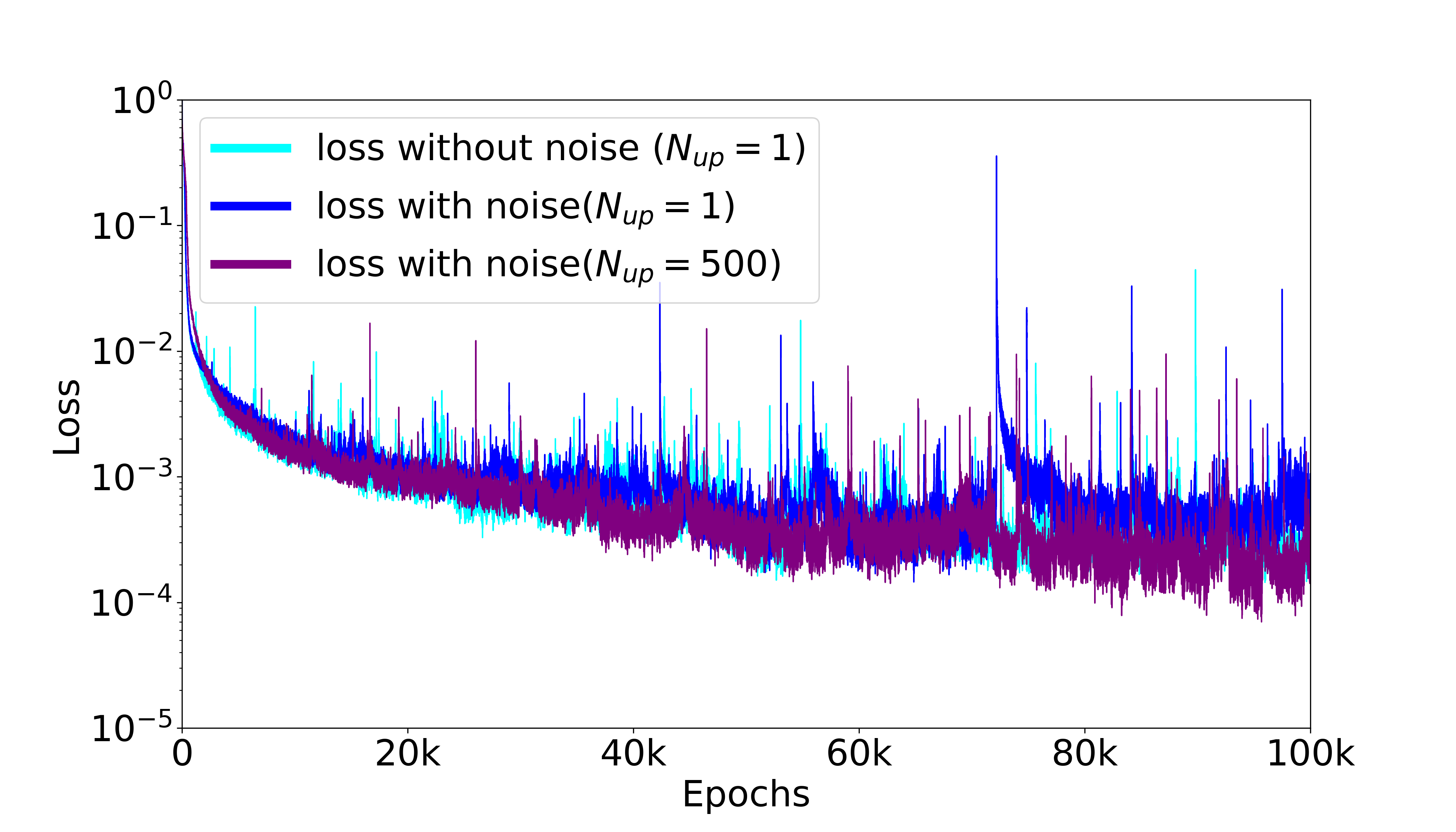 }}
    \caption{Results for CCD of planer system}
    \label{fig:2d_results}
\end{figure}

\Figref{fig:2d_results} shows results of the proposed CCD algorithm applied to the planer example.
We compare the cases where $\rho$ is sampled with and without adding the noises $\epsilon_i$ and $\epsilon_j$ at the lines 4 and 6 in Algorithm\,\ref{alg:UCCD}. 
For the additive noise, we used a uniform distribution with the amplitude of $0.1$. 
In \figref{fig:rho_nolta}, \emph{cyan} and \emph{blue} lines show the changes in the design parameter $\rho$ when it is updated at each epoch ($N_\mathrm{up}=1$). The \emph{purple} line show the results with $N_\mathrm{up}=500$ and adding noise. 
The solid curves correspond to the mean of ten repeated experiments, and the shaded region shows their standard deviations.
It can be seen that the standard deviation of $\rho$ is significantly reduced by adding the noise, and the parameter $\rho$ converges to the optimal solution $\rho = 2.1$, which is calculated in \cite{Hoshino2023}, only when $\rho$ is explored by adding the noise. 
As shown in \figref{fig:loss_nolta},  the PINN loss $L_\mathrm{PINN}$ is kept small, and it can be confirmed that the simultaneous update of $\rho$ does not impede the learning process of the PINN. 
%Overall, the proposed simultaneous update algorithm appropriately works for this example. 

%%%%%%%%%%%%%%%%%%%%%%%%%%%%%%%%%%%%%%%%%%%%%%%%%%%%%%%%%%%%%%%%%%%%%%%%
\subsection{Stochastic LQR problem} \label{sec:stochasticLQR}

Next we present a stochastic LQR example based on~\cite{Zhou2021}. 
Consider the following $d$-dimensional controlled stochastic process given by 
\begin{align}
\mathrm{d}X_t = \rho U_t \mathrm{d}t + \sqrt{2} \mathrm{d}W_t
\end{align}
where $X_t \in \mathbb{X} \subset \R^d$, $U_t \in \R^d$, $W_t \in \R^d$, and $\rho \in \R$. 
For the domain $\mathbb{X}$, consider $d$-dimensional sphere of radius $R$: 
\begin{align}
    \mathbb{X} = \{ x \in \R^d \,| \,  \| x \| < R \}. 
\end{align}
The cost functional for the optimal control is given by 
\begin{align}
    J_\mathrm{c}(\rho, x) =  \mE\biggl[ & \int_0^\tau ( p \| X_t \|^2 + q  \| U_t \|^2 -2k d) \mathrm{e}^{-\gamma s}\mathrm{d}s  \notag \\ 
    & +  \mathrm{e}^{-\gamma \tau} k R^2 \biggr] 
\end{align}
where $\tau$ stands for the exit time of the domain $\mathbb{X}$, and the constant $k$ is given by 
  $ k = (\sqrt{q^2 \gamma^2 + 4pq\rho^2} -\gamma q )/( 2 \rho^2 )$. 
With this setting, the value function $V$ satisfies the following HJB equation:
\begin{align}
 & \partial_x^2 V(\rho, x) + \inf_{u \in \R^d } (\rho u^\top \partial_x V(\rho,x) + q\|u\|^2 ) \notag \\
   &\hspace{30mm} + p\|x\|^2 -2kd -\gamma V(\rho, x) = 0.
\end{align}
with the boundary condition $V(\rho, x) = k R^2$. 
It is known that the PDE has the exact solution as a quadratic function $V(x) =k \| x^2 \| $, and the optimal control is given as $u^\ast(x)= -\rho \partial_x V(x)/(2q) = -k\rho x /q$. 
Then, the objective function for the UCCD problem is given by 
\begin{align}
 J = w_\mathrm{p} |\rho| + w_\mathrm{c} \int_\mathbb{X} \omega(x) V(x) \mathrm{d}x 
\end{align}
where $w_\mathrm{p}= w_\mathrm{c}=1$, and $\omega(x)\equiv 1/|\mathbb{X}|$.

%%%%%%%%%%%%%%%%%%%%%%%%%%%%%%%%%%%%%
% Fig: d-dimensional LQR
%%%%%%%%%%%%%%%%%%%%%%%%%%%%%%%%%%%%%
\begin{figure}[!t]
    \centering
     \subcaptionbox{Change in design parameter $\rho$\label{fig:lqr_rho}}{
     \includegraphics[width=0.98\linewidth]{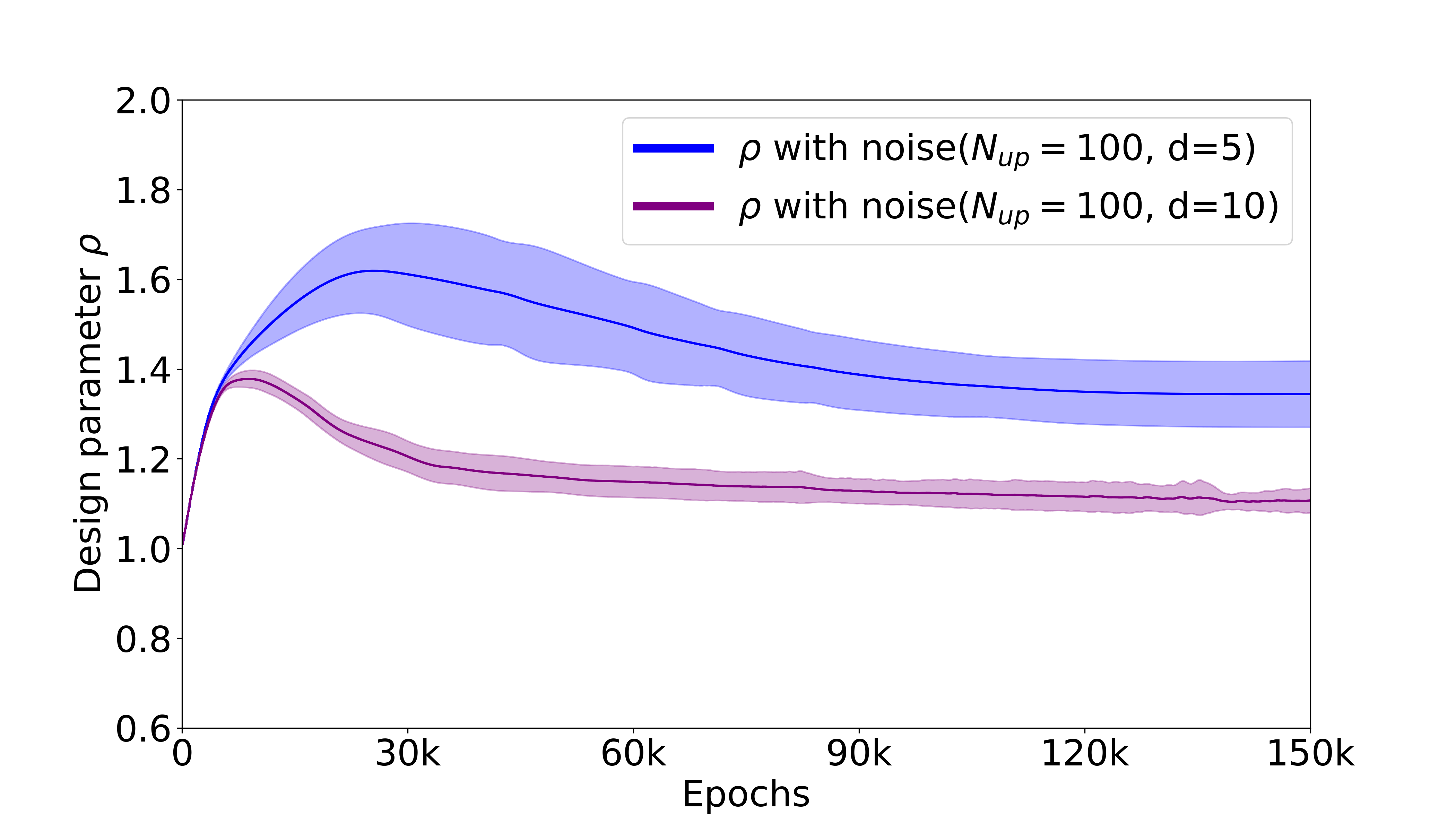}
     }\\[3mm]
     \subcaptionbox{Change in loss term $L_\mathrm{PINN}$\label{fig:lqr_loss}}{
     \includegraphics[width=0.98\linewidth]{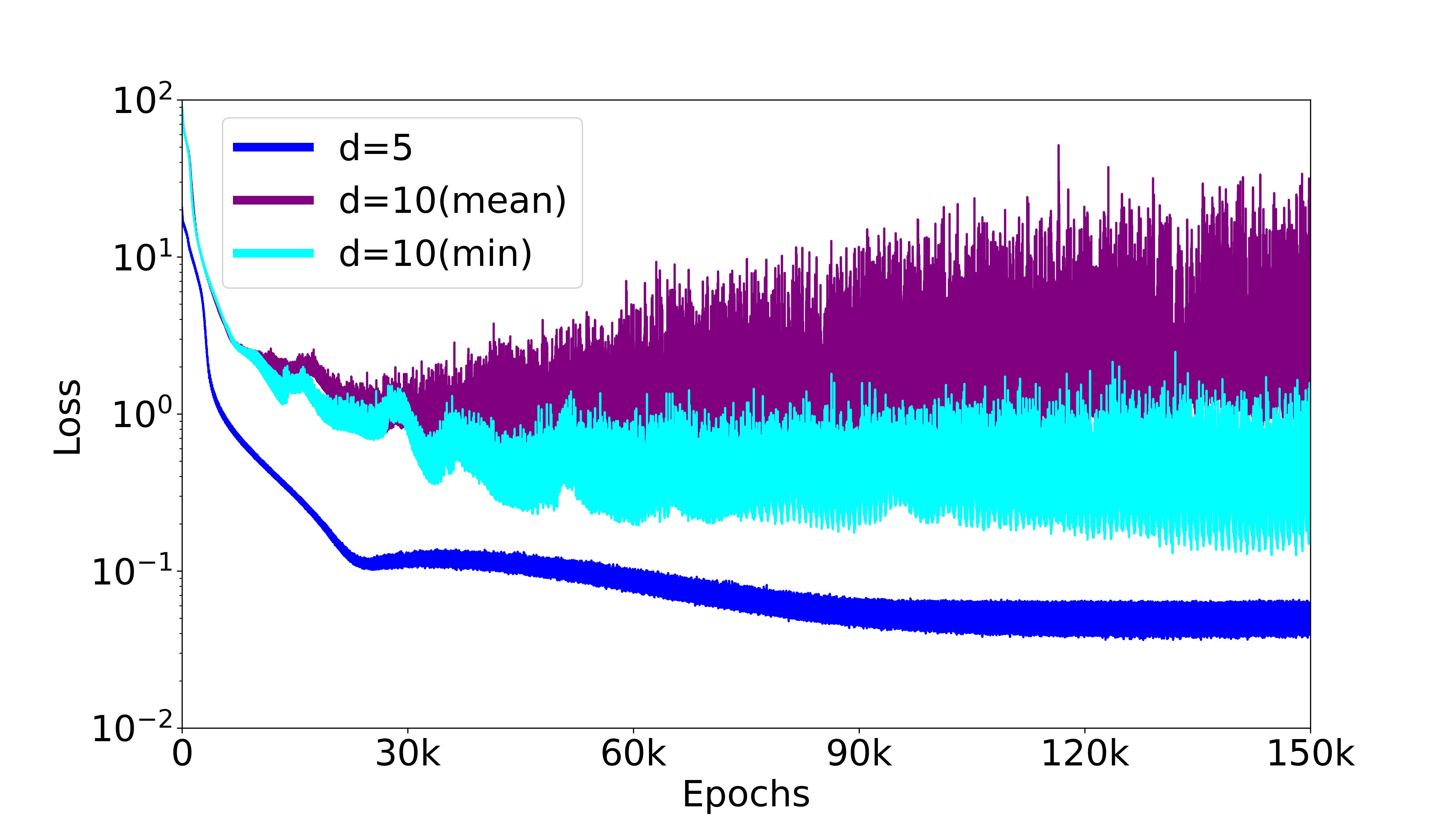 }
     }
    \caption{Results of UCCD for stochastic LQR problem}
    \label{fig:lqr_results}
\end{figure}

\Figref{fig:lqr_results} shows results for $d=5$ (shown by \emph{blue}) and $10$ (shown by \emph{purple}). The other parameters are given as $p=q=\gamma=1$, and $R=2$ and $4$ for $d=5$ and $10$, respectively. 
The shape of the neural network is the same with the above example, and the learning rates are $\SI{1e-4}{}$ and $\SI{1e-2}{}$ for $\theta$  and $\rho$, respectively, for both of $d=5$ and $10$.  
The weights in the loss \eqref{eq:loss_pinn} are $\mu_\mathrm{p}=\mu_\mathrm{b}=1$. 
The numbers of samples are $N_\mathrm{p} = N_\mathrm{b} = N_\mathrm{r} = 1000$ for $d=5$, and $N_\mathrm{p} = N_\mathrm{b} = N_\mathrm{r} = 100,000$ for $d=10$. 
\Figref{fig:lqr_rho} shows the mean (solid lines) and standard deviation (shaded area) of the design parameter $\rho$ in 10 repeated experiments, and it can be seen that $\rho$ converges at around $120,000$ epochs.
\figref{fig:lqr_loss} shows the changes in the loss function $L_\mathrm{PINN}$, and it is kept low for $d=5$ as shown by the \emph{blue} line. 
For $d=10$, \emph{purple} shows the mean value of the ten experiments, and \emph{cyan} shows the results when the loss function is the smallest among these experiments. 
Although a relatively large variation was observed in the loss, the design parameter $\rho$ converged to similar values in these 10 experiments.

%%%%%%%%%%%%%%%%%%%%%%%%%%%%%%%%%
\section{Conclusions} \label{sec:conclusion}

This paper proposed a novel UCCD problem formulation to cope with  uncertainties coming from noises acting on the dynamics and modeling errors.
The proposed method utilizes closed-loop solutions of an optimal control problem characterized by the Hamilton-Jacobi-Bellman equation as a PDE constraint in the UCCD problem.  
 A solution algorithm is developed based on Physics-informed Neural Networks (PINNs), and numerical examples show that simultaneous update of PINN weights and the design parameters effectively works for solving UCCD problems.

Future directions of this work include extension of the proposed algorithm to address problems where the optimal control is obtained only in an implicit form. 
A possible approach for this task is to use a reinforcement learning framework to obtain an optimal policy. 
Another future direction is to used the proposed method in practical applications in \emph{e.g.}, robotics and energy systems as mentioned in \secref{sec:introduction}.

%%%%%%%%%%%%%%%%%%%%%%%%%%%%%%%%%%%%%%%
\section*{Acknowledgment}

This work was supported in part by JST, ACT-X Grant Number JPMJAX210M, Japan, and the Kansai Research Foundation for Technology Promotion, Japan.

%%%%%%%%%%%%%%%%%%%%%%%%%%%%%%%%%%%%%%%%%%%%%
\bibliography{scis_isis2024_codesign}
\bibliographystyle{IEEEtran}

% \bibitem{b1} G. Eason, B. Noble, and I. N. Sneddon, ``On certain integrals of Lipschitz-Hankel type involving products of Bessel functions,'' Phil. Trans. Roy. Soc. London, vol. A247, pp. 529--551, April 1955.
% \bibitem{b2} J. Clerk Maxwell, A Treatise on Electricity and Magnetism, 3rd ed., vol. 2. Oxford: Clarendon, 1892, pp.68--73.
% \bibitem{b3} I. S. Jacobs and C. P. Bean, ``Fine particles, thin films and exchange anisotropy,'' in Magnetism, vol. III, G. T. Rado and H. Suhl, Eds. New York: Academic, 1963, pp. 271--350.
% \bibitem{b4} K. Elissa, ``Title of paper if known,'' unpublished.
% \bibitem{b5} R. Nicole, ``Title of paper with only first word capitalized,'' J. Name Stand. Abbrev., in press.
% \bibitem{b6} Y. Yorozu, M. Hirano, K. Oka, and Y. Tagawa, ``Electron spectroscopy studies on magneto-optical media and plastic substrate interface,'' IEEE Transl. J. Magn. Japan, vol. 2, pp. 740--741, August 1987 [Digests 9th Annual Conf. Magnetics Japan, p. 301, 1982].
% \bibitem{b7} M. Young, The Technical Writer's Handbook. Mill Valley, CA: University Science, 1989.
% \end{thebibliography}

\end{document}